# Excitation of coherent optical phonons in iron garnet by femtosecond laser pulses

Pritam Khan[1,2*], Masataka Kanamaru[1], Wei-Hung Hsu[1], Minori Kichise[1], Yasuhiro Fujii[3], Akitoshi Koreeda[3], and Takuya Satoh[1]

[1]Department of Physics, Kyushu University, Fukuoka 819–0395, Japan

[2]Department of Physics & Bernal Institute, University of Limerick, V94 T9PX Limerick, Ireland

[3]Department of Physical Sciences, Ritsumeikan University, Kusatsu 525–8577, Japan

E-mail: khan@email.phys.kyushu-u.ac.jp, Pritam.Khan@ul.ie

**ABSTRACT**: We employed femtosecond pump–probe technique to investigate the dynamics of coherent optical phonons in iron garnet. A phenomenological symmetry-based consideration reveals that oscillations of the terahertz $T_{2g}$ mode are excited. Selective excitation by a linearly polarized pump and detection by a circularly polarized probe confirm that impulsive stimulated Raman scattering (ISRS) is the driving force for the coherent phonons. Experimental results obtained from ISRS measurements reveal excellent agreement with spontaneous Raman spectroscopy data, analyzed by considering the symmetry of the phonon modes and corresponding excitation and detection selection rules.

# 1. Introduction

Excitation and detection of coherent phonons have attracted significant scientific interest over the past few decades, as they can be employed to investigate the dynamic properties of various materials, from metals to dielectrics.[1-11] Impulsive stimulated Raman scattering (ISRS) provides a unique approach to study the dynamics of coherent phonons through inelastic scattering of a single femtosecond laser pulse.[12] ISRS can be used for the selective excitation of coherent phonons by choosing the proper polarization of the pump pulses.[13-15]

Garnet is an important class of material because of its unique properties, which are desirable for applications in fabricating waveguide isolators,[16] microwave filters,[17] magneto-optical devices,[18] etc. Several experimental and theoretical analyses have been performed to identify the phonon modes in garnet by means of spontaneous Raman spectroscopy.[19-24] All of the spontaneous Raman studies reveal the existence of $A_{1g}$, $E_g$, and $T_{2g}$ phonon modes in the THz regime. Nevertheless, there are no experiments on the coherent optical phonon dynamics in garnet for excitations with femtosecond laser pulses, in spite of several studies on acoustic phonon dynamics.[25,26] Such studies are of importance to reveal new insights into the ultrafast processes that are triggered by femtosecond laser pulses.

In this study, we investigated the excitation and detection mechanisms of coherent optical phonons in rare-earth iron garnet by employing femtosecond pump–probe technique. Experimental results reveal the excitation of coherent phonons with a frequency of 4.2 THz, which is consistent with the $T_{2g}$ mode. By comparing the results of the pump–probe measurements with the spontaneous Raman scattering and by considering the very low absorption coefficient of the garnet with respect to pump pulses, we confirm that ISRS is responsible for the excitation of coherent phonons.

## 2. Methods

*2.1 Sample preparation*

We used a single crystal of $Gd_{1.5}Yb_{0.5}BiFe_5O_{12}$ (GdYbBIG) with a (111)-plane orientation, grown by the liquid-phase epitaxy method.[27] GdYbBIG is a paraelectric single crystal with cubic symmetry *m3m* with no structural multidomains. The lateral dimensions of the sample were 5 mm × 5 mm, while its thickness was $d$ = 400 μm. The *x, y,* and *z* axes of the sample are defined to be parallel to the $[11\bar{2}]$, $[1\bar{1}0]$, and [111] crystallographic axes, respectively,[28] which are represented by an illustration of the crystal orientation in Fig. 1(a). The sample is a ferrimagnetic insulator, with a Curie temperature of 573 K[29] and compensation temperature of 96 K.[18] It is important to note that, GdYbBIG is a popular garnet as an optical isolator and owing to the large Faraday rotation. Its spin wave from GHz to THz range has been studied extensively before.[28-33] Therefore, it is quite natural to explore the possibility of exciting the coherent phonon modes in the same sample which leads to the motivation of the present study.

*2.2 Femtosecond pump-probe spectroscopy*

To excite and detect the coherent phonons in the GdYbBIG single crystal by femtosecond laser pulses, we performed time-resolved pump–probe measurements in transmission geometry. Pump pulses with a duration of 70 fs and wavelength of 1300 nm were generated from a Ti:sapphire laser coupled with a regenerative amplifier and optical parametric amplifier [34] (OPA, Spectra-Physics) at a repetition rate of 500 Hz. In our experiment, we used a conventional commercial TOPAS OPA [35] which is a state-of-the-art instrument for wavelength extension of the Spitfire Ti:Sapphire amplifier system. The TOPAS is entirely computer controlled which minimizes the time adjustment of the laser system and maximizes the experimental productivity.

The pump beam was incident at an angle of 10° with respect to the sample normal. Probe pulses with a wavelength of 800 nm, duration of 50 fs and repetition rate of 1 kHz were directed perpendicular to the sample. The linearly polarized pump beam was focused on a spot diameter of 85 μm. The probe beam was circularly polarized, with a spot size two times smaller than that of the pump beam. The power of the probe beam was kept at 60 μW. In our experiments, optically excited coherent phonons were observed using the polarimetric detection (PD) technique with balanced detectors.[4,6,11,15, 36–38]. This detection scheme can extract the contributions of an induced change in the ellipticity of the probe polarization. A schematic diagram of the pump–probe set up is shown in Fig. 1(b). We have also denoted $\theta$ and $\psi$ in Fig. 1(a) to describe the pump azimuthal angle and the detection angle, respectively. The pump azimuthal angle $\theta$ is defined as the angle between the direction of the pump polarization and the *x* axis, whereas $\psi$ determines the angle of the Wollaston prism, used in the detection scheme and therefore termed as detection angle. In all pump–probe measurements, we applied an in-plane magnetic field of 1.5 kOe so that the sample remained in the monodomain state.

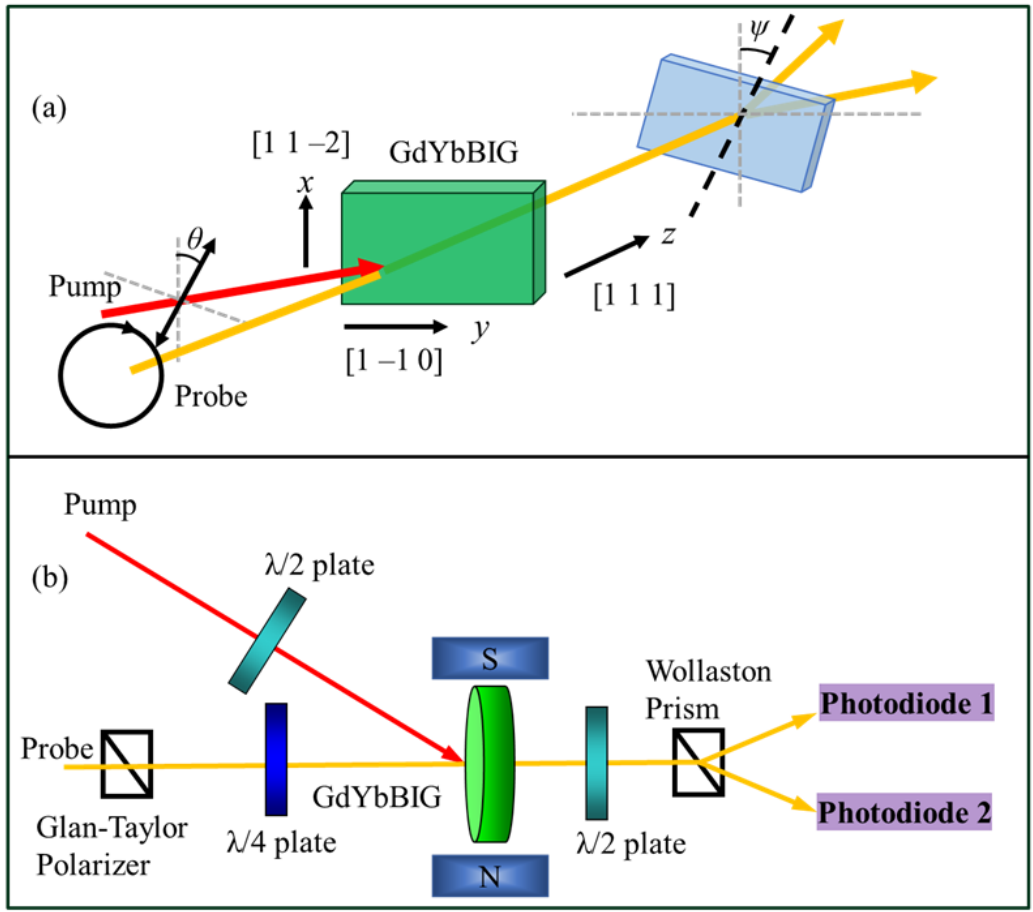

**Figure 1**. (a) Crystal orientation in GdYbBIG. (b) The experimental set up for pump–probe measurements.

*2.3 Spontaneous Raman spectroscopy*

To identify the symmetry of the phonon modes in GdYbBIG, we measured the polarized spontaneous Raman spectra in the backscattering geometry. We used a 785 nm laser as the excitation light source in both linear and circular polarized configurations. The power of the laser was kept at 45 mW. The laser spot diameter on the sample was 20 μm. Our homemade Raman setup is equipped with a spectrometer (Acton SpectraPro-2500i) with a ruled grating of 1200 gr/mm and a multichannel detector (Hamamatsu S7031-1007S). The spectral resolution of our Raman setup is 60 GHz. We performed spontaneous Raman measurements in four different geometric configurations, as illustrated in Fig. 2. The polarization of the scattered light was selected using an analyzer.[39] For linearly polarized light, when the polarizations of the scattered light are parallel (HH) and perpendicular (HV) to that of the incident light, the scattered Raman intensities are denoted $I_{HH}$ and $I_{HV}$, respectively. For the right-handed circularly polarized incident light, when the scattered lights are right-handed (RR) and left-handed (RL) circularly polarized, they are denoted $I_{RR}$ and $I_{RL}$, respectively. Here, right- and left-handed circularly polarized light are defined by the projection of light polarization onto the sample plane[40], independent of the light travelling direction. To make the Raman data consistent with pump–probe measurements, we also applied an in-plane magnetic field of 2.3 kOe.

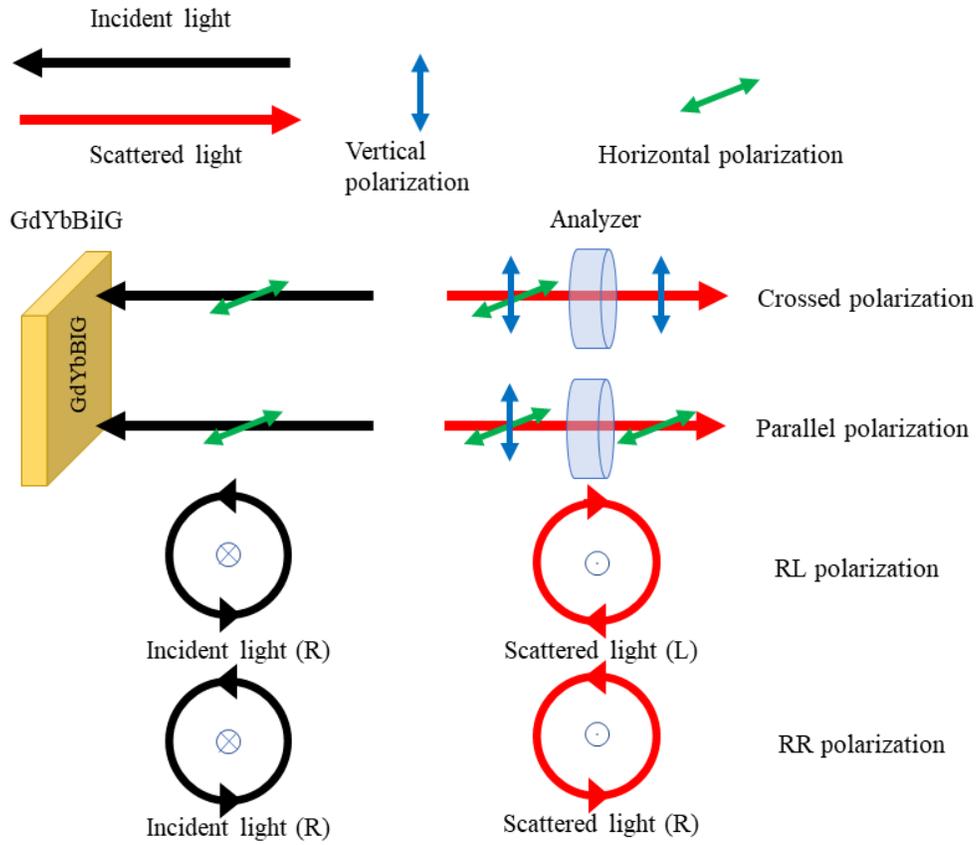

**Figure 2**. The spontaneous Raman spectroscopy setup in HH, HV, RR, and RL polarization configurations.

## Results and Discussion

Figure 3(a) demonstrates the time-resolved ellipticity changes ($\Delta\eta$) in GdYbBIG as a function of the probe delay ($t$) at $T = 300$ K for an excitation with $\theta = \pi/2$ and detection with $\psi = \pi/4$. The pump fluence was fixed at 60 mJ/cm$^2$. We observed a strong coherent artifact at the pump–probe overlap ($t = 0$), followed by a pronounced oscillatory signal. In general, coherent artifact is induced by third-order nonlinear interaction between the pump and probe beams at zero-time delay.[41,42] We made use of this coherent artifact to find the $t = 0$ position. The oscillation is relaxed over a few picoseconds; however, an intense abrupt signal similar to that at $t = 0$ is observed at $t \sim 6.2$ ps.

When $t \sim 6.2$ ps, a part of the pump pulse is reflected from the second face of the sample, counter-propagates, and is reflected again from the front face; it then overlaps with the probe pulse in the sample, yielding an intense abrupt signal. The sample surface is perpendicular to the probe, so it is tilted by 10° with respect to the pump. Therefore, the optical length travelled by the pump is $d_{\text{pu}} = 406$ μm. Using the refractive index of the pump ($n_{\text{pu}} = 2.3$), we calculated that the abrupt signal appears at

$$t = \frac{2 d_{\text{pu}} n_{\text{pu}}}{c_0}, \quad (1)$$

where $c_0$ is the velocity of light in vacuum. Equation (1) shows that $t \sim 6.2$ ps, which is in good agreement with the experimental results of Fig. 3(a).

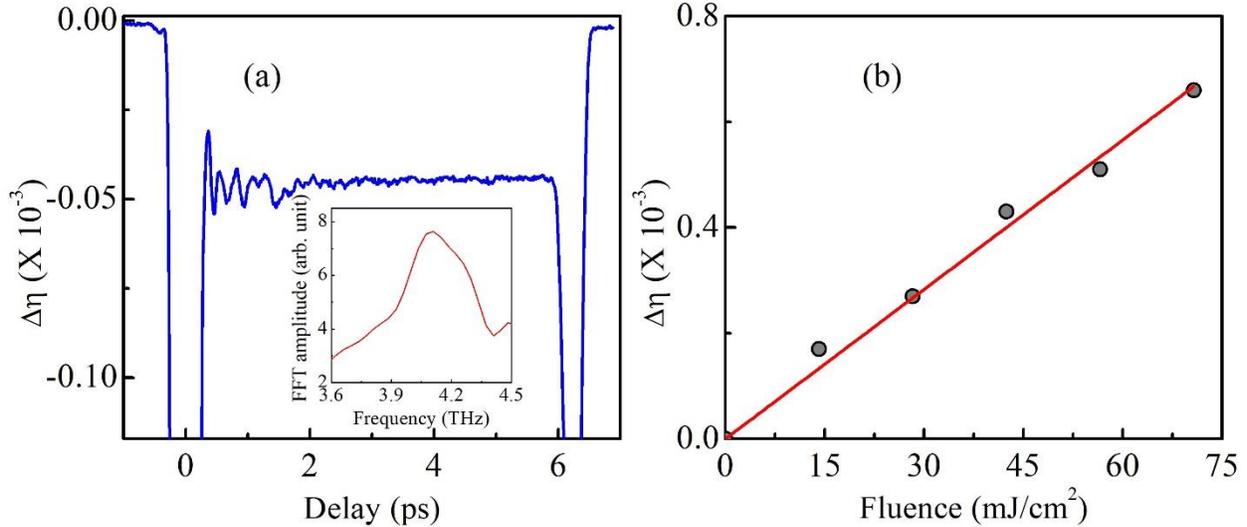

**Figure 3**. (a) Temporal evolution of the ellipticity changes in the transmitted probe polarization of GdYbBIG at 300 K. The inset represents the FFT amplitude spectrum of the temporal oscillation, revealing the center frequency of 4.2 THz. (b) Variation of amplitude of 4.2-THz mode as a function of the pump fluence.

To quantify the oscillatory signal, we performed a fast Fourier transform (FFT) of the waveform at positive delays, which yielded the amplitude spectrum, as shown in the inset in Fig. 3(a). The FFT spectrum reveals that the center frequency of the oscillation is 4.2 THz. Figure 3(b) reveals the amplitude of the 4.2-THz oscillation as a function of pump fluence. Δη reveals a linear relationship with pump fluence below 70 mJ/cm$^2$. Clearly, 60 mJ/cm$^2$ falls in the linear regime, and we can therefore confirm that our measurements are well within the limit of the ISRS process.

Next, we performed pump–probe measurements at various temperatures (80–300 K) with fixed $\theta = \pi/2$ and $\psi = \pi/4$, as shown in Fig. 4(a). The 4.2-THz oscillation dominates the signal; however, near the compensation temperatures, e.g., at 80 and 120 K, we observed a low-frequency oscillation of a few hundred gigahertz. This low-frequency gigahertz oscillation is most likely associated with the periodic Faraday rotation caused by the residual perpendicular component of the magnetization.[43] Notably, the abrupt signal is always observed at $t \sim 6.2$ ps, independent of the measuring temperature.

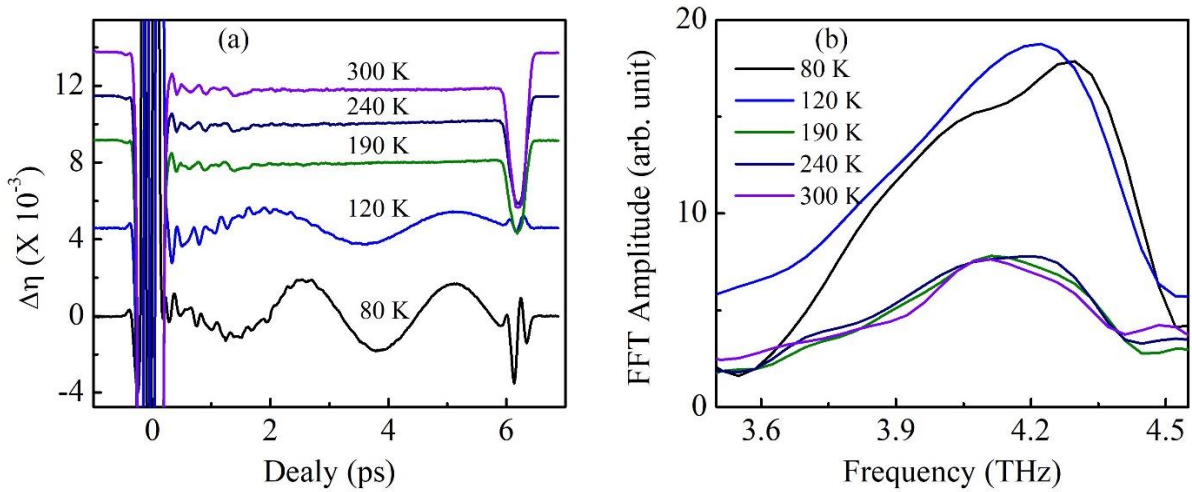

**Figure. 4**. (a) Time-resolved ellipticity changes at various temperatures. (b) FFT amplitude spectra of time-resolved ellipticity curves at various temperatures.

Figure 4(b) represents the FFT amplitude spectra of the oscillation waveforms at the considered temperatures. The 4.2-THz mode does not change its peak position in the temperature range 80–300 K. Such observation clearly reveals that the 4.2-THz mode is associated with the excitation of phonons, instead of magnons.

To determine the symmetry of the 4.2-THz phonon modes that are excited in the pump–probe measurements, we recorded the spontaneous Raman spectrum of GdYbBIG at 300 K in HH, HV, RL, and RR polarization configurations, as demonstrated in Fig. 5(a). We observed a center peak at 4.1 THz in all polarization configurations. The strong peak at 4.1 THz matches well with that of the pump–probe measurements. Therefore, we can assume that the peaks found in spontaneous Raman and pump–probe experiments can be assigned to the same symmetry modes. Considering that the garnet crystal belongs to the cubic *m3m* point group, the possible symmetry modes are $A_{1g}$, $E_g$, or $T_{2g}$. The Raman tensors of the phonon modes in our *x, y,* and *z* coordinates can be expressed as in TABLE I: [44]

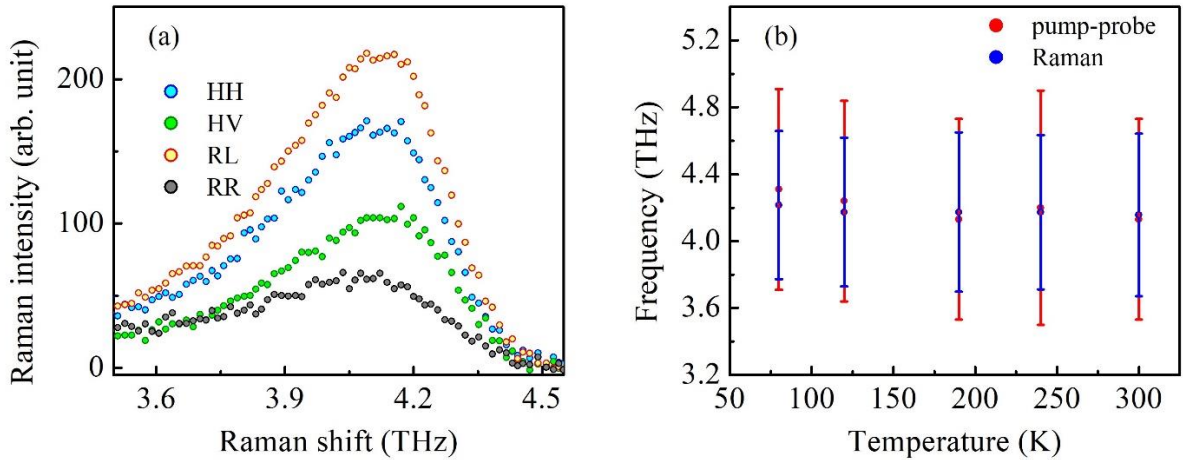

**Figure 5.** Spontaneous Raman spectrum of GdYbBIG at 300 K in HH, HV, RR, and RL polarization configurations. A peak is observed at a frequency of 4.1 THz. (b) Comparison of the central frequency obtained from pump–probe and spontaneous Raman measurements at various measuring temperatures. Here the red and blue lines represent the bars which is determined by the line width of the phonon mode in pump-probe and Raman measurements, respectively.

| $A_{1g}$ | $E_g(1)$ | $E_g(2)$ | $T_{2g}(1)$ | $T_{2g}(2)$ | $T_{2g}(3)$ |
|---|---|---|---|---|---|
| $\begin{pmatrix} d & 0 & 0 \\ 0 & d & 0 \\ 0 & 0 & d \end{pmatrix}$ | $\begin{pmatrix} -b & 0 & \sqrt{2}b \\ 0 & b & 0 \\ \sqrt{2}b & 0 & 0 \end{pmatrix}$ | $\begin{pmatrix} 0 & b & 0 \\ b & 0 & \sqrt{2}b \\ 0 & \sqrt{2}b & 0 \end{pmatrix}$ | $\begin{pmatrix} 2a & 0 & 2\sqrt{2}a \\ 0 & -6a & 0 \\ 2\sqrt{2}a & 0 & 4a \end{pmatrix}$ | $\begin{pmatrix} -4a & 2\sqrt{3}a & -\sqrt{2}a \\ 2\sqrt{3}a & 0 & -\sqrt{6}a \\ -\sqrt{2}a & -\sqrt{6}a & 4a \end{pmatrix}$ | $\begin{pmatrix} -4a & -2\sqrt{3}a & -\sqrt{2}a \\ -2\sqrt{3}a & 0 & \sqrt{6}a \\ -\sqrt{2}a & \sqrt{6}a & 4a \end{pmatrix}$ |

**TABLE I**. Raman tensor of the phonon modes in GdYbBIG in our *x, y,* and *z* coordinates, associated with $A_{1g}$, $E_g$, and $T_{2g}$ symmetry modes.

To uniquely identify the symmetry of the phonon modes, we compare the Raman scattering intensities associated with the $A_{1g}$, $E_g$, and $T_{2g}$ phonons in the HH, HV, RR, and RL configurations. The calculated and experimental ratio $I_{HH}$ : $I_{HV}$ : $I_{RR}$ : $I_{RL}$ is listed in TABLE II:

| $A_{1g}$ | $E_g$ | $T_{2g}$ | Experiment |
|---|---|---|---|
| 1: 0: 1: 0 | 1: 1: 0: 2 | 3: 2: 1: 4 | 2.7: 1.7: 1: 3.55 |

**TABLE II**. Comparison of Raman scattering intensities in GdYbBIG, calculated for $A_{1g}$, $E_g$, and $T_{2g}$ symmetry modes, with those obtained in the experiment.

From Fig. 5(a), we obtained $I_{HH}$ = 166, $I_{HV}$ = 104, $I_{RR}$ = 61, and $I_{RL}$ = 217 at 4.1 THz, which gives $I_{HH}$ : $I_{VV}$ : $I_{RR}$ : $I_{RL}$ = 2.7: 1.7: 1: 3.55. Therefore, our results indicate that the 4.1-THz mode observed in the spontaneous Raman spectrum of garnet is consistent with the $T_{2g}$ phonon mode.

In Fig. 5(b), the central frequency (4.1 THz) of the $T_{2g}$ mode at various temperatures obtained by the Raman measurements are compared with those obtained from the pump–probe measurements. These frequencies are consistent over a wide temperature range within the bar. Here the bars in both measurements are determined by the line width of the phonon mode. The temperature dependence of these frequencies does not exhibit a noticeable change, which indicates that anharmonic phonon-phonon coupling is negligibly small in the sample. We conjecture that the anharmonicity might be obscured by the inhomogeneous broadening of the Raman spectra of the GdYbBIG, where $Gd^{3+}$, $Yb^{3+}$, and $Bi^{3+}$ ions are randomly located in the dodecahedral positions in the garnet structure. Apart from phonon-phonon coupling, the electron-phonon interaction is also not observed because photon energy of the pump pulse is below the bandgap, so carriers are not excited in GdYbBIG.

After finding that the 4.1-THz mode is consistent with the $T_{2g}$ phonon mode and understanding its temperature dependence, it is important to identify the coherent excitation mechanisms for this phonon mode with respect to femtosecond laser pulses. Importantly, the sample is transparent to the pump wavelength (the absorption coefficient is 0.3 cm$^{-1}$ for 1300 nm). Additionally, in our present study, the linearly polarized pump excites the asymmetric $T_{2g}$ mode. Therefore, the displacive excitation of coherent phonons (DECP) process resulting from direct excitation, which is associated with the excitation of symmetric $A_g$ modes in opaque materials,[45-48] can be excluded. If the excitation mechanism is based on the ISRS, then the distinct polarization of the pump provides the most efficient conditions for the excitation of coherent phonons. First, we need to calculate the driving force of the coherent phonon modes, which depends on the electric field of the pump pulses. Since the pump light is propagating in the *z* direction, we can express the pump polarization as $(e_x, e_y, e_z) = (\cos\theta, \sin\theta, 0)$. For transparent media, the classical equation of motion of a lattice field is expressed as:[13,49]

$$\frac{d^2Q}{dt^2} + \Omega^2 Q = F(t). \tag{2}$$

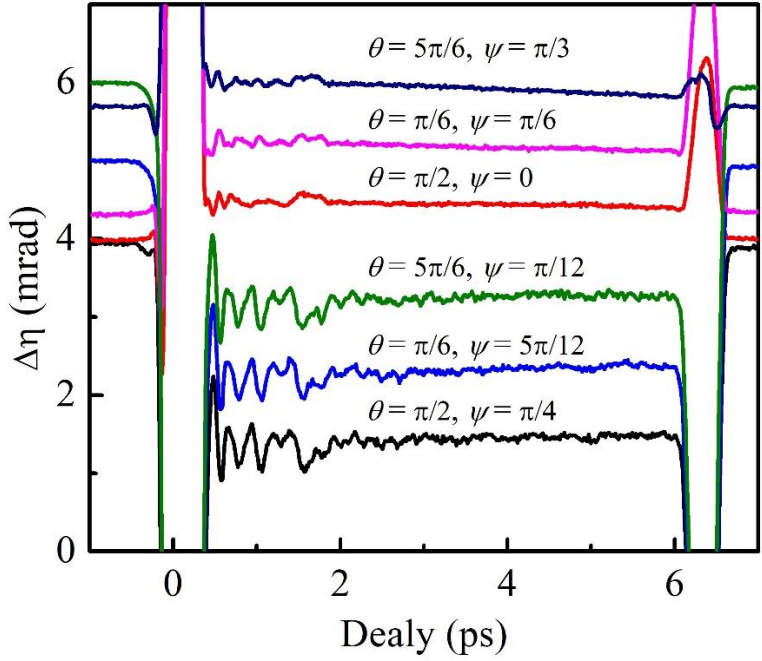

**Figure 6.** Time-resolved ellipticity changes in the transmitted probe polarization of GdYbBIG at several pump azimuthal angles and detection angles.

Here, $Q$ is the normal coordinate of phonon with angular frequency $\Omega$. The driving force $F$ is often modeled through ISRS and given by $F \propto R_{ij} e_i e_j^*$, where $R_{ij}$ is the Raman tensor of the sample. Consequently, the driving force of the $T_{2g}$ phonon mode can be expressed as:

$$F[T_{2g}(1)] \propto \mathrm{Re}[2a e_x e_x^* - 6a e_y e_y^*] = 2a[2\cos(2\theta) - 1], \tag{3}$$

$$F[T_{2g}(2)] \propto \mathrm{Re}[-4a e_x e_x^* + 2\sqrt{3}a(e_x e_y^* + e_y e_x^*)] = 2a[2\cos(2\theta - 2\pi/3) - 1], \tag{4}$$

$$F[T_{2g}(3)] \propto \mathrm{Re}[-4a e_x e_x^* - 2\sqrt{3}a(e_x e_y^* + e_y e_x^*)] = 2a[2\cos(2\theta + 2\pi/3) - 1], \tag{5}$$

The Raman tensor matrix and driving force calculation provides a unique way to selectively excite the $T_{2g}(1)$, $T_{2g}(2)$, and $T_{2g}(3)$ modes. We found from Eqs. (3)–(5) that, when $\theta = \pi/2$, $5\pi/6$, and $\pi/6$, we can selectively excite only $T_{2g}(1)$, $T_{2g}(2)$, and $T_{2g}(3)$ modes, respectively.

Next, we move on to focus on the detection scheme. Let us consider the following eigenequation:

$$\widehat{N}^2 e - \tilde{\epsilon} e = 0, \tag{6}$$

where $\widehat{N}$ and $\tilde{\epsilon}$ are the refractive index and dielectric tensor, used in the eigenequation, respectively. After applying Eq. (6) to the Raman tensors of $T_{2g}(1)$, $T_{2g}(2)$, and $T_{2g}(3)$ modes, we find the determinants of Eq. (6) takes the following forms respectively:

$$T_{2g}(1): \begin{vmatrix} \widehat{N}^2 - (\epsilon + 2a) & 0 \\ 0 & \widehat{N}^2 - (\epsilon - 6a) \end{vmatrix} = 0 \tag{7}$$

$$T_{2g}(2): \begin{vmatrix} \widehat{N}^2 - (\epsilon - 4a) & -2\sqrt{3}a \\ -2\sqrt{3}a & \widehat{N}^2 - \epsilon \end{vmatrix} = 0 \tag{8}$$

$$T_{2g}(3): \begin{vmatrix} \widehat{N}^2 - (\epsilon - 4a) & 2\sqrt{3}a \\ 2\sqrt{3}a & \widehat{N}^2 - \epsilon \end{vmatrix} = 0 \tag{9}$$

Here $\epsilon$ is the dielectric constant which is unmodulated by the contribution from $T_{2g}$ phonon. By substituting the eigenvalue $\widehat{N}$ obtained from Eqs. (7)–(9) in Eq. (6), we obtain the eigenpolarizations as:

$$T_{2g}(1): \begin{pmatrix} e_x \\ e_y \end{pmatrix} = \begin{pmatrix} 1 \\ 0 \end{pmatrix} \text{ and } \begin{pmatrix} 0 \\ 1 \end{pmatrix} \tag{10}$$

$$T_{2g}(2): \begin{pmatrix} e_x \\ e_y \end{pmatrix} = \frac{1}{2}\begin{pmatrix} 1 \\ \sqrt{3} \end{pmatrix} \text{ and } \frac{1}{2}\begin{pmatrix} -\sqrt{3} \\ 1 \end{pmatrix} \tag{11}$$

$$T_{2g}(3): \begin{pmatrix} e_x \\ e_y \end{pmatrix} = \frac{1}{2}\begin{pmatrix} 1 \\ -\sqrt{3} \end{pmatrix} \text{ and } \frac{1}{2}\begin{pmatrix} \sqrt{3} \\ 1 \end{pmatrix} \tag{12}$$

for $\widehat{N} = N_\alpha$ and $N_\beta$, respectively, where $N_\alpha = \sqrt{\epsilon + 2a}$ and $N_\beta = \sqrt{\epsilon - 6a}$. Consequently, for linearly polarized light, the azimuthal angles of the eigenpolarization are 0 and $\pi/2$ for $T_{2g}(1)$, $\pi/3$ and $-\pi/6$ for $T_{2g}(2)$, and $-\pi/3$ and $\pi/6$ for $T_{2g}(3)$, for $N_\alpha$ and $N_\beta$ respectively. Let us consider the case when one of these modes is excited. The circularly polarized probe before entering the sample related to this mode in the directions with unit vectors $\boldsymbol{\alpha}$ and $\boldsymbol{\beta}$, can be expressed as:

$$e_{\text{pr}} = \frac{1}{\sqrt{2}} \begin{pmatrix} 1 \\ -i \end{pmatrix} \qquad (13)$$

Since there is difference of refractive index $\widehat{N}$ in Eqs. (10)–(12), the polarization of the probe pulses after transmitting through the sample of thickness $d$ can be expressed as:

$$e'_{\text{pr}} = \frac{1}{\sqrt{2}} \begin{pmatrix} \exp(-i\frac{\omega N_\alpha}{c}d) \\ -i \exp(-i\frac{\omega N_\beta}{c}d) \end{pmatrix} \qquad (14)$$

Since there is a phase difference between the two eigenpolarizations, the circularly polarized probe becomes elliptic with an axis at an intermediate angle between the two orthogonal eigenpolarizations. Therefore, if the detection angle $\psi$ is set parallel to the eigenpolarization, we would not be able to detect any phonon modes. For maximum efficient detection, $\psi$ should be angled at $\pm \pi/4$ with respect to eigenpolarization. In this regard, TABLE III represents the selection rules for the excitation and detection of $T_{2g}$ mode based on the previous discussion.

| Mode | Excitation ($\theta$) | Detection max ($\psi$) | Detection min ($\psi$) |
|---|---|---|---|
| $T_{2g}(1)$ | $\pi/2$ | (i) $\pi/4$ | (ii) 0 |
| $T_{2g}(2)$ | $5\pi/6$ | (iii) $\pi/12$ | (iv) $\pi/3$ |
| $T_{2g}(3)$ | $\pi/6$ | (v) $5\pi/12$ | (vi) $\pi/6$ |

**TABLE III**. Selection rule for the excitation and detection of the $T_{2g}$ phonon modes.

Figure 6 exemplifies the temporal evolution of the ellipticity curves of the probe polarization following the six combinations of $\theta$ and $\psi$. The ellipticity curves reveal the oscillation of frequency 4.2 THz, corresponding to the combination of (i), (iii), and (v) as predicted in TABLE III.

Conversely, the ellipticity curves do not exhibit remarkable oscillation for (ii), (iv), and (vi). We believe that the small non-zero oscillation observed in the latter case is due to the angular deviation of the polarizing wave plates and the Wollaston prism etc.

In conclusion, we performed a comprehensive study on the excitation and detection of coherent phonons in iron garnet by pump–probe PD measurements. We analytically calculated the driving force for the phonon in the garnet by considering its *m3m* point-group symmetry. Our calculations predict the specific experimental conditions for the excitation and detection of coherent phonons, depending on the pump and probe polarizations. Experimental results indicated that linearly polarized laser pulses excited coherent phonons with a frequency of 4.2 THz by ISRS. By comparing the results obtained from the pump–probe measurements with the spontaneous Raman spectra, and considering the symmetry of the phonon modes in garnet, we found that the 4.2-THz mode is consistent with the $T_{2g}$ phonon mode. Furthermore, observation of optical coherent phonon modes opens up the possibility of studying magnon-phonon coupling in GdYbBIG.

## Acknowledgements

This study was supported by Japan Society for the Promotion of Science (JSPS) KAKENHI (numbers JP15H05454, JP16F16358, JP17K18765, and JP26103004) and JSPS Core-to-Core Program (A. Advanced Research Networks).